\documentclass[11pt,a4paper]{article}
\usepackage{jheppub_kim}
\usepackage{longtable}

\usepackage{epsfig}
\usepackage{amsmath}
\usepackage{graphicx}
 \usepackage{graphics}
 \usepackage[scriptsize,nooneline,hang]{caption}
\usepackage[hang,nooneline,scriptsize]{subfigure}

\usepackage{array}
\usepackage{makecell}
\begin{document}

\title{Accretion disks around a static black hole in $f(R)$ gravity}

\author[a]{Saheb Soroushfar}
\affiliation[a]{Faculty of Technology and Mining, Yasouj University, Choram 75761-59836, Iran}
\author[b,c]{and Sudhaker Upadhyay}
\affiliation[b]{Department of Physics, K.L.S. College, Nawada-805110 (A Constituent Unit of Magadh University, Bodh-Gaya), India }
\affiliation[c]{Visiting Associate, Inter-University Centre for Astronomy and Astrophysics (IUCAA) Pune, Maharashtra-411007}
 
\emailAdd{soroush@yu.ac.ir} 
\emailAdd{sudhakerupadhyay@gmail.com}

\abstract{We provide a description of a thin accretion disc for a static spherically symmetric black holes in $f(R)$ gravity. In this regard, we  first study the horizons of black holes in $f(R)$ gravity. The equation of motion and effective potential are also computed which eventually leads to  possible existence of innermost circular orbits of accretion disc. 
We derive the specific energy,   specific angular momentum and  
angular velocity  of the particles moving in circular orbits also. A comparative study of  various parameters are also presented.  The locations of the event horizon, cosmological horizon,    innermost and outermost stable circular orbits are also pointed out. 
}

\maketitle

\section{Introduction}  
Since the last decades,  people are trying  to solve the the problems like the accelerated expansion of the universe \cite{1,2} and the dark matter origin \cite{3}  with the help of  modified theories of gravity.  There are various modified theory of gravity. $f(R)$ theory of gravity  is one of them \cite{6}. This is because general relativity (GR) agrees to gravity
with high accuracy  when the curvature is small \cite{4}, however for very large values of the
curvature there does not exist any such evidence.
 In this regard,   black holes  
are the ideal places to look for modified  GR \cite{5}. The features of static and spherically symmetric black hole solutions  in
$f(R)$  gravity theories are discussed \cite{7}.   
Recently, a linear
perturbations of a Kerr black hole in $f(R)$ gravity using the Newman-Penrose formalism
are  studied in Ref. \cite{8}. 

It is expected  that, at large distances,  the geometry of the space-time in $f(R)$ modified gravity models
must be different from that of standard GR. Therefore, one needs to develop a
method that could   observationally distinguish  and test  the possible deviations from GR. The study of
accretion disks around compact objects could be one of the possible methods.
 It is well-known that  most of the astrophysical objects like black holes grow   in mass due to accretion.
Some observations suggest that  around black
holes, there exist gas clouds  together
with an associated accretion disc  \cite{9}. The first comprehensive study of mass accretion around rotating black holes  in GR was made in \cite{10}.
After that general relativistic
model of thin accretion disk was developed  under the assumption that the disk is in a
steady-state  (i.e. the mass accretion rate  is constant in
time and does not depend of the radius of the disk) \cite{11,12,13}. 
The radiation properties of the thin accretion disks  had been given  in
\cite{14,15}.
In fact, the descriptions of the accretion disk around
  wormholes, non-rotating/rotating quark, boson/fermion stars, brane world
black holes, naked singularities     and
$f (R)$ modified gravity models  have been given in Refs. \cite{16,17,18,19,20,21,22,23,24,25,26,27,28,29,30,31,32,33}.

In order to study the parameters specifying the thin accretion disc for the static spherically symmetric black hole in $f(R)$ gravity, we first write the metric and the action for the system. From the metric function, we analyze the horizon of the black hole.  Here we obtain  that existence of event horizon varies according to   different dimensionless parameters.  For 
instance,  for certain values of these parameters   there exist  and  inner  black hole event horizon and outer cosmological horizon; in spite of that in certain cases   there exists
only single black hole event horizon. In certain cases, the event and cosmological
horizons collapse   and   naked singularities occur. Furthermore,   we evaluate equations of motion and effective potential of the system where we note that energy and the angular momentum together with the
cosmological constant and a real parameter describe the shape of the orbit. 

It is well-known that the accretion disk around  the black hole appears due to the particles revolving in a circular orbit. Therefore, we compute the specific energy, the specific angular momentum and the
angular velocity  of the particles moving in circular orbits for the  the static spherically symmetric black hole in $f(R)$ gravity.
From  the necessary condition for existence of innermost circular orbits of accretion disc
we obtain an expression for dimensionless cosmological constant $ \tilde{\Lambda} $  as a function of the radial coordinate of the circular orbits.   Here, we do comparative analysis
of $ \tilde{\Lambda} $ corresponding to different values of dimensionless real parameter and 
find that the value of  dimensionless cosmological constant  increases with larger values of real parameter. We also analyze 
  $ \tilde{\Lambda} $ as a function of the radial co-ordinate of the event horizon and  as a function of the radial coordinate of the innermost stable circular orbits.
Moreover, we  plot  the effective potential  with the location of innermost circular orbit where the locations of the innermost stable circular orbits are indicated.  We provide  two tables specifying   
locations of the event and cosmological horizons, and of the innermost and outermost stable circular orbits in a static black hole in $f(R)$ gravity  and locations of the event horizon, and of the innermost stable circular orbit in a static black hole in $f(R)$ gravity, respectively.

 The paper is presented as following. In section \ref{field}, we
 discuss the structure of horizons of the static
spherically symmetric black hole in $f(R)$ gravity and derive equations of motion together with effective potential. The specific energy, specific angular momentum  and the
angular velocity 
  of the particles moving in circular orbits are calculated in section \ref{acc}.
  Finally, we conclude our work with future remarks in section \ref{section6}. 
\section{The structure of spacetime in $f(R)$ gravity}\label{field}
In this section, we recapitulate metric structure of the static
spherically symmetric black hole in $f(R)$ gravity.
Let us begin by writing a generic action for $f(R)$ gravity in four dimensional spacetime as 
\begin{equation}\label{action}
S=\frac{1}{2k}\int d^{4}{x}\sqrt{-g}f(R)+S_{m},
\end{equation}
where $k$ is Einstein's constant, $R$ is the Ricci scalar and $S_{m}$ is the matter part.
The variation of above action with respect to the metric leads to  the following  field equations:
\begin{equation}
f'(R)R_{\mu\nu}-\frac{1}{2}f(R)g_{\mu\nu}-(\nabla_{\mu}\nabla_{\nu}-g_{\mu\nu}\square)f'(R)=kT_{\mu\nu},
\end{equation}
in which $ \square=\nabla_{\alpha}\nabla^{\alpha}$ and $ f'(R)=\frac{df(R)}{dR} $.
The line element representing a 4-dimensional static black hole in $f(R)$ gravity is given by \cite{Saffari:2007zt}
\begin{equation}\label{metric}
ds^{2}=-A(r)dt^{2}+A(r)^{-1}dr^{2}+r^{2}(d\theta^{2}+\sin^{2}\theta d\varphi^{2}),  
\end{equation}
with metric function
\begin{equation}
g(r)= 1-\frac{2 M}{r}+\beta{r}-\frac{1}{3}\Lambda{r}^{2},
\end{equation}
where  $  M$ is the mass, $\beta$ is a real constant and $\Lambda$ is the cosmological constant \cite{Saffari:2007zt,Soroushfar:2015wqa}.
Here one should mention that non-zero cosmological constant are crucial 
for various phenomenon  of static and spherically symmetric black holes \cite{a,b}.
 The notion of static radius is crucial here as it can represent a natural boundary of gravitationally bound systems in an expanding universe governed by a cosmological constant, as clearly demonstrated in a variety of situations \cite{c,d,e,f,g,h}.
\subsection{The Structure of Horizons} 
The horizon of the spacetime given by equation (\ref{metric}) can be determined by setting  the condition $ g_{00}(r)=0 $, i.e.
\begin{equation}\label{Horizon}
1-\frac{2 M}{r}+\beta r-\frac{\Lambda}{3} r^{2}=0 .
\end{equation}
Now, we define following dimensionless  quantities \cite{Soroushfar:2015wqa}:
\begin{equation}\label{tilde}
\tilde{r}=\frac{r}{M},\qquad \tilde{\Lambda}=\Lambda M^{2},\qquad \tilde{\beta}=\beta M ,  
\end{equation}
which leads to the equation (\ref{Horizon}) to the following form:
\begin{equation}\label{Horizon2}
\tilde{\Lambda} \tilde{r}^{3}-\tilde{\beta} \tilde{r}^{2}-\tilde{r}+2=0 .
\end{equation}
\begin{figure}[h]
	\centering
	\subfigure{
		\includegraphics[width=0.6\textwidth]{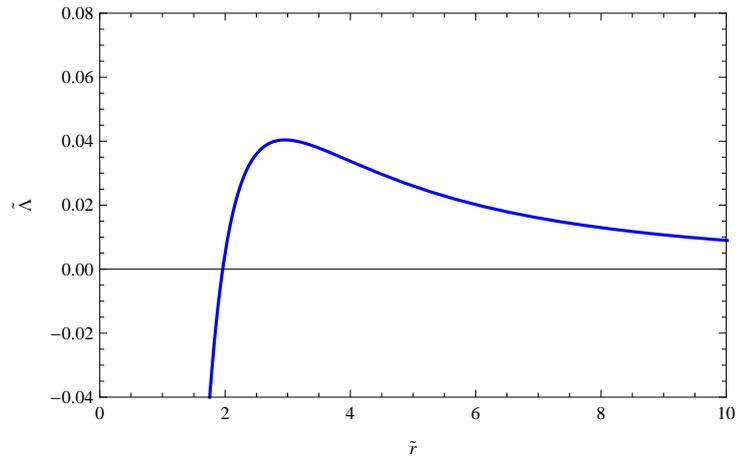}
	}
	\caption{$\tilde{\Lambda}$ vs. $\tilde{r}$ for $\tilde\beta=0.01$  }
	\label{Pic:lambda1}
\end{figure}
From Fig. \ref{Pic:lambda1}, corresponding to $\tilde\beta=0.01$,  there exist an inner black hole event horizon and an
outer cosmological horizon for
 $\tilde{\Lambda}\in (0,  0.04069)$. However,  there exists
 only single black hole event horizon for $\tilde{\Lambda} \leq 0$.
  The event and cosmological
horizons collapse for $\tilde{\Lambda} = 0.04069$ and for higher values naked singularities occur and hence, no black
holes are possible. Consequently,  the trajectories should be studied for $\tilde{\Lambda}\in (-\infty,  0.04069]$.
\begin{figure}[h]
	\centering
	\subfigure{
		\includegraphics[width=0.6\textwidth]{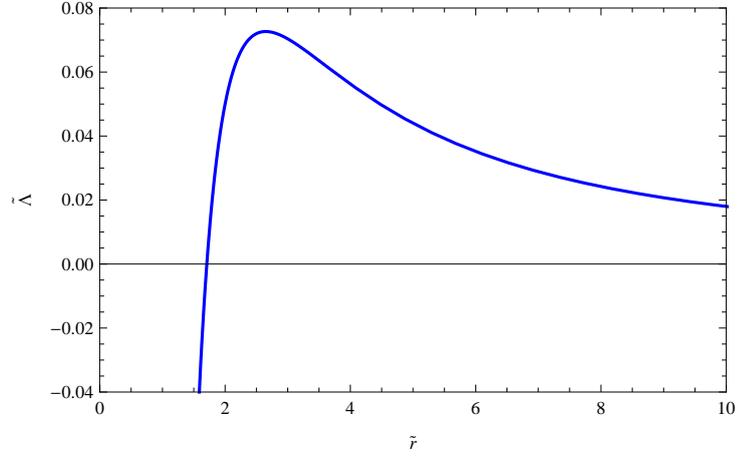}
	}
	\caption{$\tilde{\Lambda}$ vs. $\tilde{r}$ for $\tilde\beta=0.1$  }
	\label{Pic:lambda2}
\end{figure}

From Fig. \ref{Pic:lambda2}, corresponding to $\tilde\beta=0.1$,  there exist an inner black hole event horizon and an 
outer cosmological horizon for
 $\tilde{\Lambda}\in (0,  0.07226)$. However,  there exists
 only single black hole event horizon for $\tilde{\Lambda} \leq 0$.
  The event and cosmological
horizons collapse for $\tilde{\Lambda} = 0.07226$ and for higher values naked singularities occur and hence, no black
holes are possible. Consequently,  the trajectories should be studied for $\tilde{\Lambda}\in (-\infty,  0.07226]$.
\begin{figure}[h]
	\centering
	\subfigure{
		\includegraphics[width=0.6\textwidth]{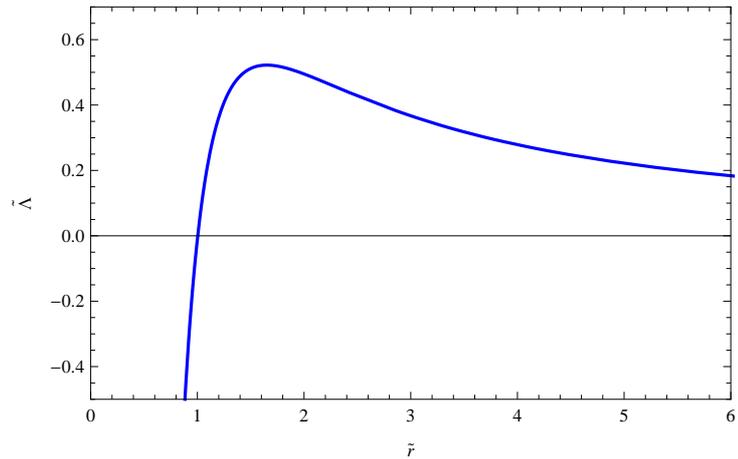}
	}
	\caption{$\tilde{\Lambda}$ vs. $\tilde{r}$ for $\tilde\beta=0.99$  }
	\label{Pic:lambda3}
\end{figure}

From Fig. \ref{Pic:lambda3}, corresponding to $\tilde\beta=0.99$,  there exist an inner black hole event horizon and an 
outer cosmological horizon for
 $\tilde{\Lambda}\in (0,  0.5259)$. However,  there exists
 only single black hole event horizon for $\tilde{\Lambda} \leq 0$.
  The event and cosmological
horizons collapse for $\tilde{\Lambda} = 0.5259$ and for higher values naked singularities occur and hence, no black
holes are possible. Consequently,  the trajectories should be studied for $\tilde{\Lambda}\in (-\infty,  0.5259]$.
\begin{figure}[h]
	\centering
	\subfigure[]{
		\includegraphics[width=0.4\textwidth]{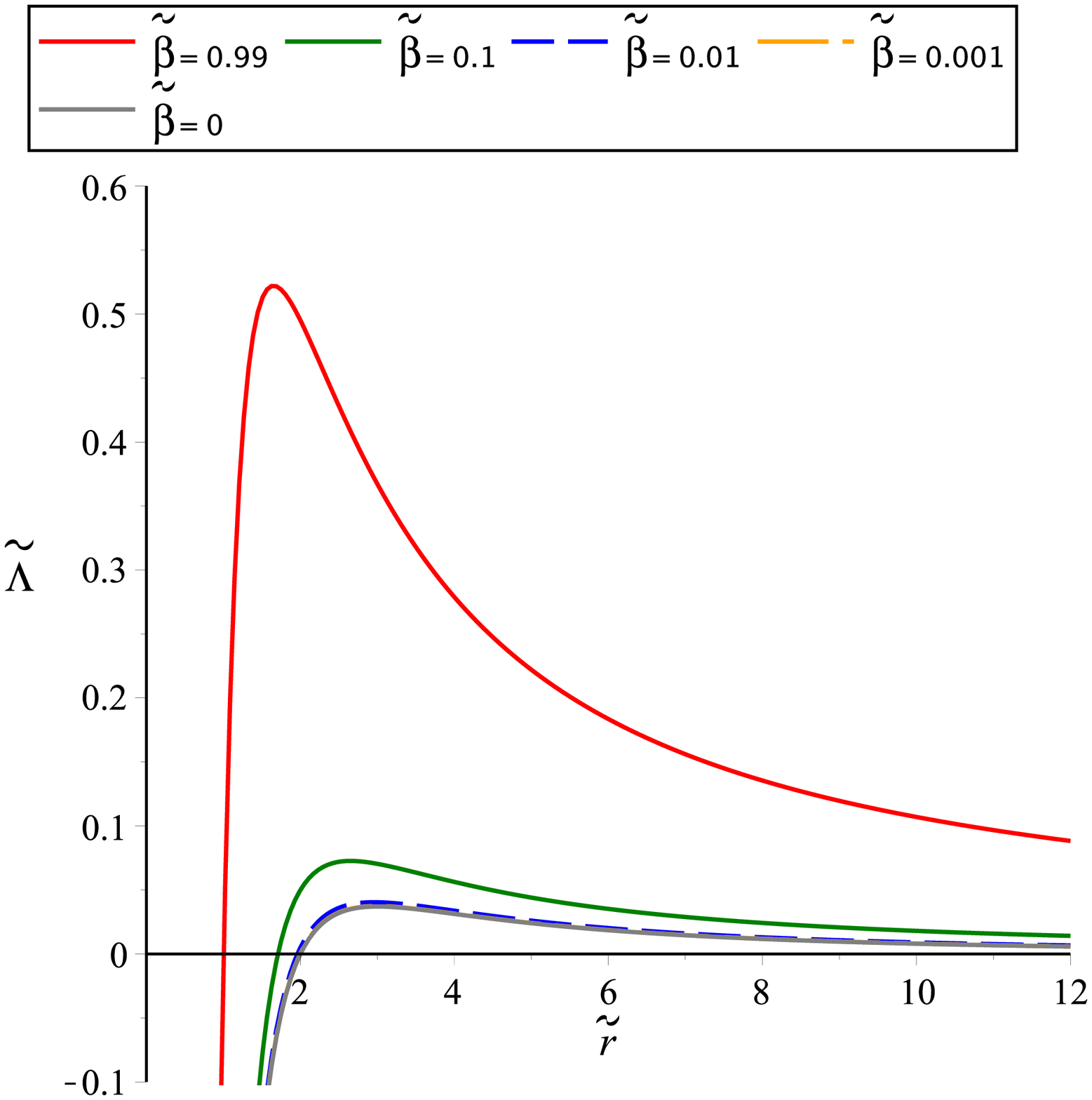}
	}
		\subfigure[Closeup of (a)]{
			\includegraphics[width=0.4\textwidth]{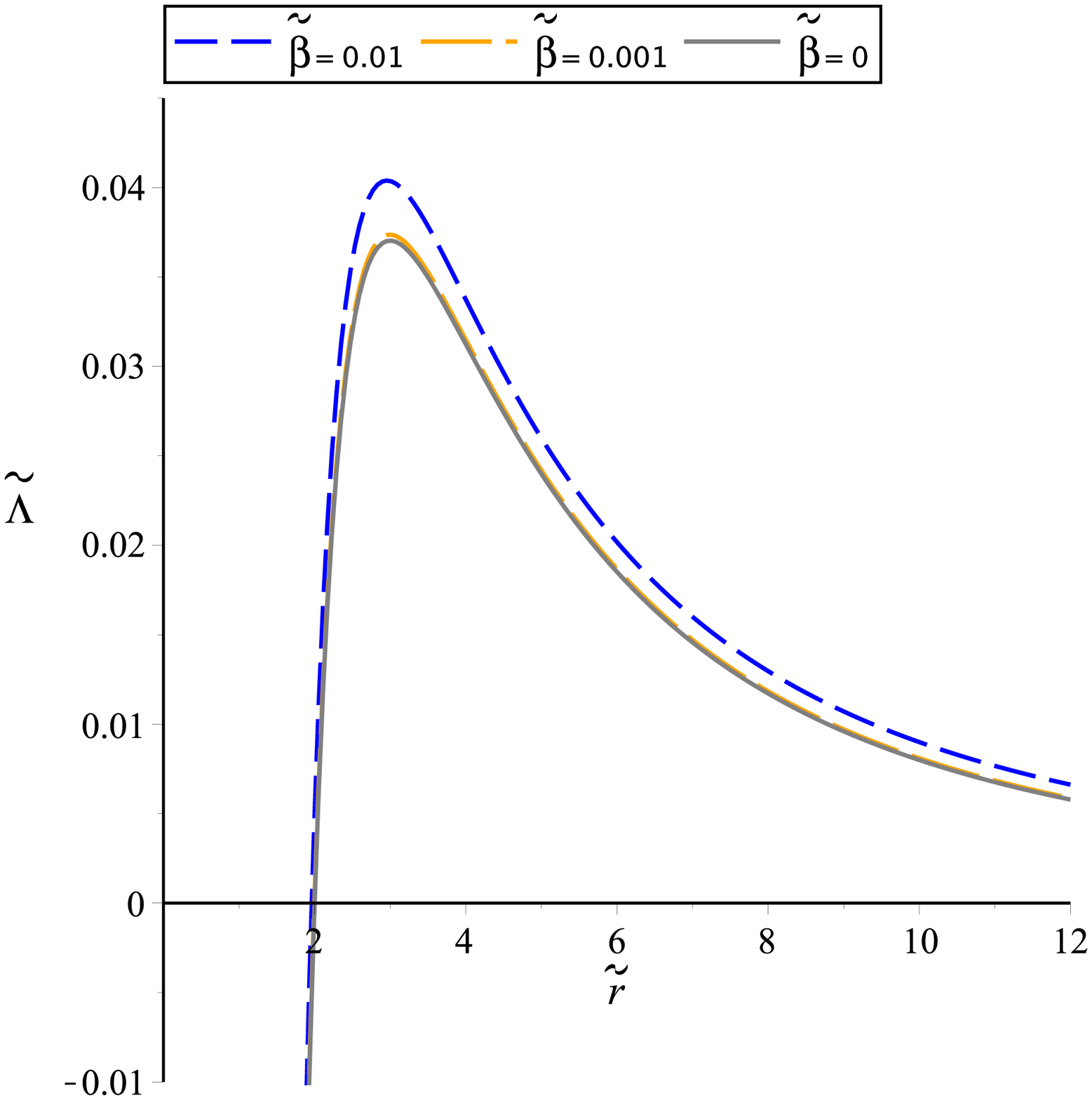}
		}
	\caption{Plot of $\tilde{\Lambda}$ as a function of the radial coordinate of the event horizon for different values of $\tilde\beta$. }
	\label{Pic:LambdaBetaH}
\end{figure}

A comparative analysis of these plots are given in Fig. \ref{Pic:LambdaBetaH} which reflects that as long as the values of $\tilde{\beta}$ reduces the peak value of $\tilde{\Lambda}$ 
becomes smaller.

\subsection{Equations of motion and effective potential}

In order to study the dynamics of the system, we evaluate equations of motion and effective potential. In order to do so, we first write the Lagrangian $ \mathcal{L} $ for a point particle in the space–time characterized by metric (\ref{metric})  as:
\begin{equation}
\mathcal{L}=\frac{1}{2}g_{\mu\nu}\frac{dx^\mu}{ds}\frac{dx^\nu}{ds}=\frac{1}{2}\epsilon.
\end{equation}
Here, $\epsilon$ takes value $1$ for massive particle and $0$ for photon (massless).
In the equatorial plane, conserved energy $E$ and angular momentum $L$ read
\begin{eqnarray}
E&=&g_{tt}\frac{dt}{ds}=\left(1-\frac{2m}{r}+\beta{r}-\frac{1}{3}\Lambda{r}^{2}\right)\frac{dt}{ds},\\
L&=&g_{\varphi\varphi}\frac{d\varphi}{ds}=r^{2}\frac{d\varphi}{ds}.
\end{eqnarray}
By considering above equations, the geodesic equations corresponding to massive particles can be obtained as \cite{Soroushfar:2015wqa}
\begin{eqnarray}
\left(\frac{dr}{ds}\right)^2&=&E^2-\left(1-\frac{2m}{r}+\beta{r}-\frac{1}{3}\Lambda{r}^{2}\right)\left(1+\frac{L^2}{r^2}\right),\label{dr/ds}\\ 
\left(\frac{dr}{d\varphi}\right)^2&=&\frac{r^4}{L^2}\left[E^2-\left(1-\frac{2m}{r}+\beta{r}-\frac{1}{3}\Lambda{r}^{2}\right)\left(1+\frac{L^2}{r^2}\right)\right], \label{dr/dphi}\\
 \left(\frac{dr}{dt}\right)^2&=&\frac{1}{E^2}\left(1-\frac{2m}{r}+\beta{r}-\frac{1}{3}\Lambda{r}^{2}\right)^2\left[E^2-\left(1-\frac{2m}{r}+\beta{r}-\frac{1}{3}\Lambda{r}^{2}\right)\left(1+\frac{L^2}{r^2}\right)\right].\label{dr/dt}
\end{eqnarray} 
Eqs. (\ref{dr/ds}), (\ref{dr/dphi}) and (\ref{dr/dt}) provide us a complete description of the
dynamics. However, Eq. (\ref{dr/ds}) reflects the expression of an effective
potential
\begin{equation}\label{V}
V_{eff}=\left(1-\frac{2m}{r}+\beta{r}-\frac{1}{3}\Lambda{r}^{2}\right)\left(1+\frac{L^2}{r^2}\right).
\end{equation}
Using Eqs. (\ref{tilde})  and (\ref{V}), we have
\begin{equation}\label{Vtilde}
V_{eff}=\left(1-\frac{2}{\tilde{r}}+\tilde\beta{\tilde{r}}-\frac{1}{3}\tilde\Lambda{\tilde{r}}^{2}\right)\left(1+\frac{\tilde{L}^2}{\tilde{r}^2}\right),
\end{equation}
where $\tilde{L}=L/M$.
Here, we note that   energy  and the angular momentum    together with the cosmological constant  and $\beta$ describe the shape of the orbit.

\section{Thin accretion disk}\label{acc}
\begin{figure}[h] 
	\centering
	\subfigure[]{
		\includegraphics[width=0.4\textwidth]{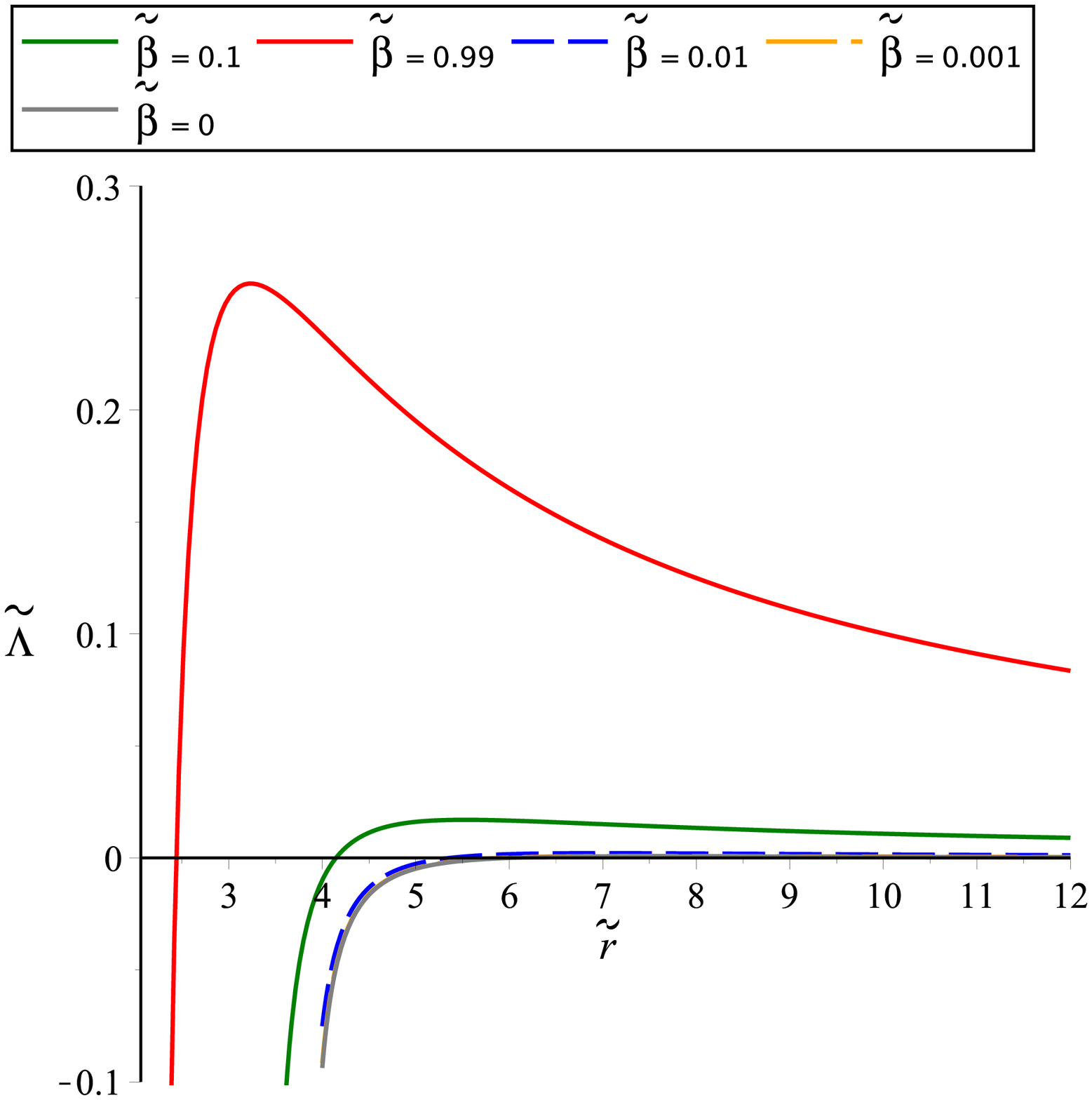}
	}
	\subfigure[Closeup of (a)]{
		\includegraphics[width=0.4\textwidth]{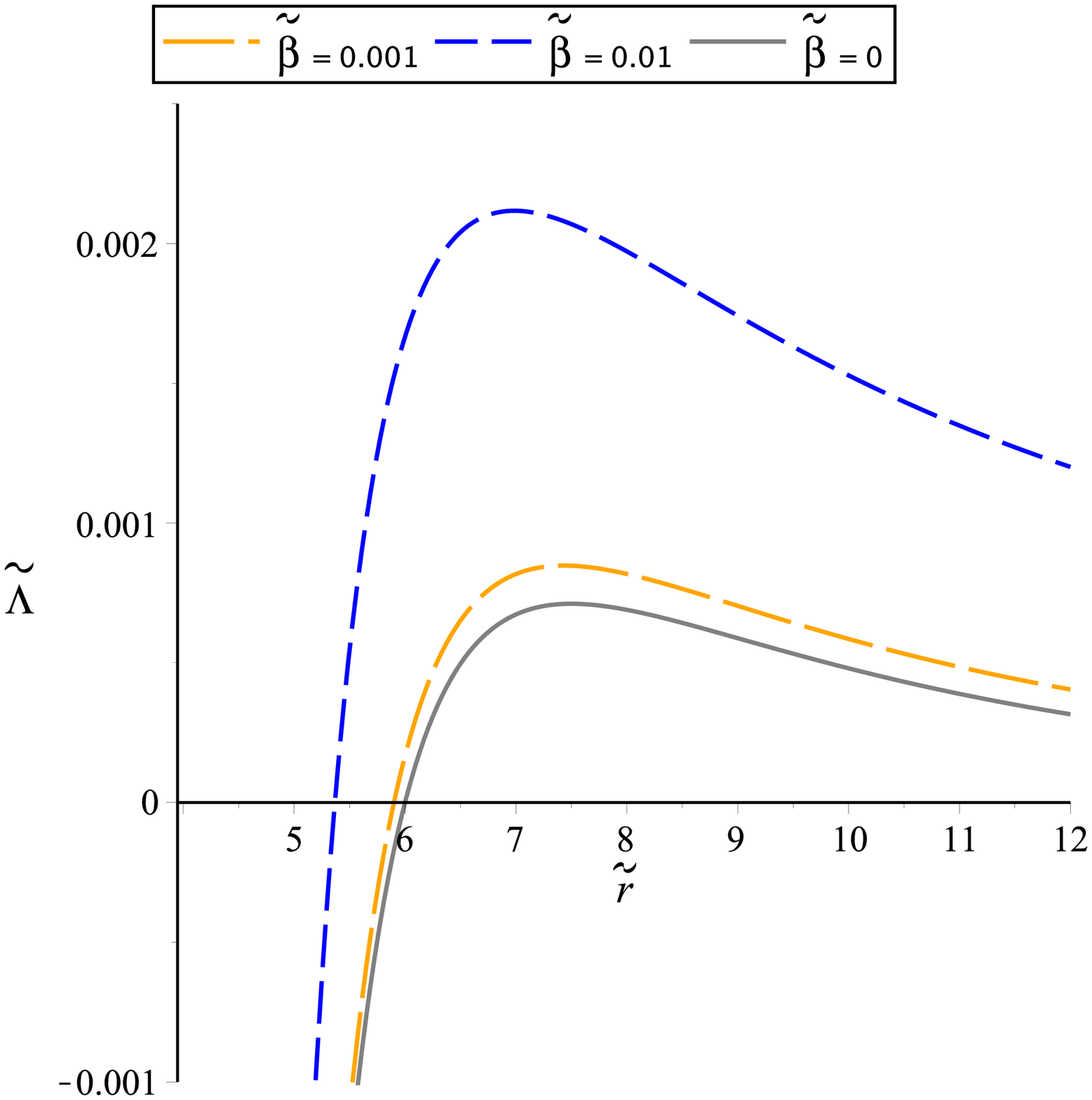}
	}
	\caption{Plot of $\tilde{\Lambda}$ as a function of the radial coordinate of the innermost stable circular orbits for different values of $\tilde\beta$. }
	\label{Pic:LambdaBeta}
\end{figure}

\begin{figure}[h]
	\centering
	\subfigure{
		\includegraphics[width=0.5\textwidth]{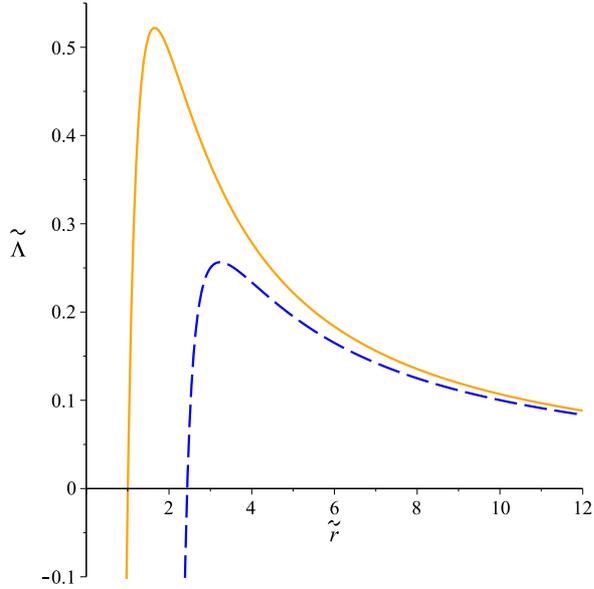}
	}
	\caption{Plot of $\tilde{\Lambda}$ as a function of the radial coordinate of the event horizon (orange line) and $\tilde{\Lambda}$ as a function of the radial coordinate of the innermost stable circular orbits (blue dashed line) for $\tilde\beta=0.99$.}
	\label{Pic:LambdaBeta12}
\end{figure}

\begin{figure}[h]
	\centering
	\subfigure{
		\includegraphics[width=0.6\textwidth]{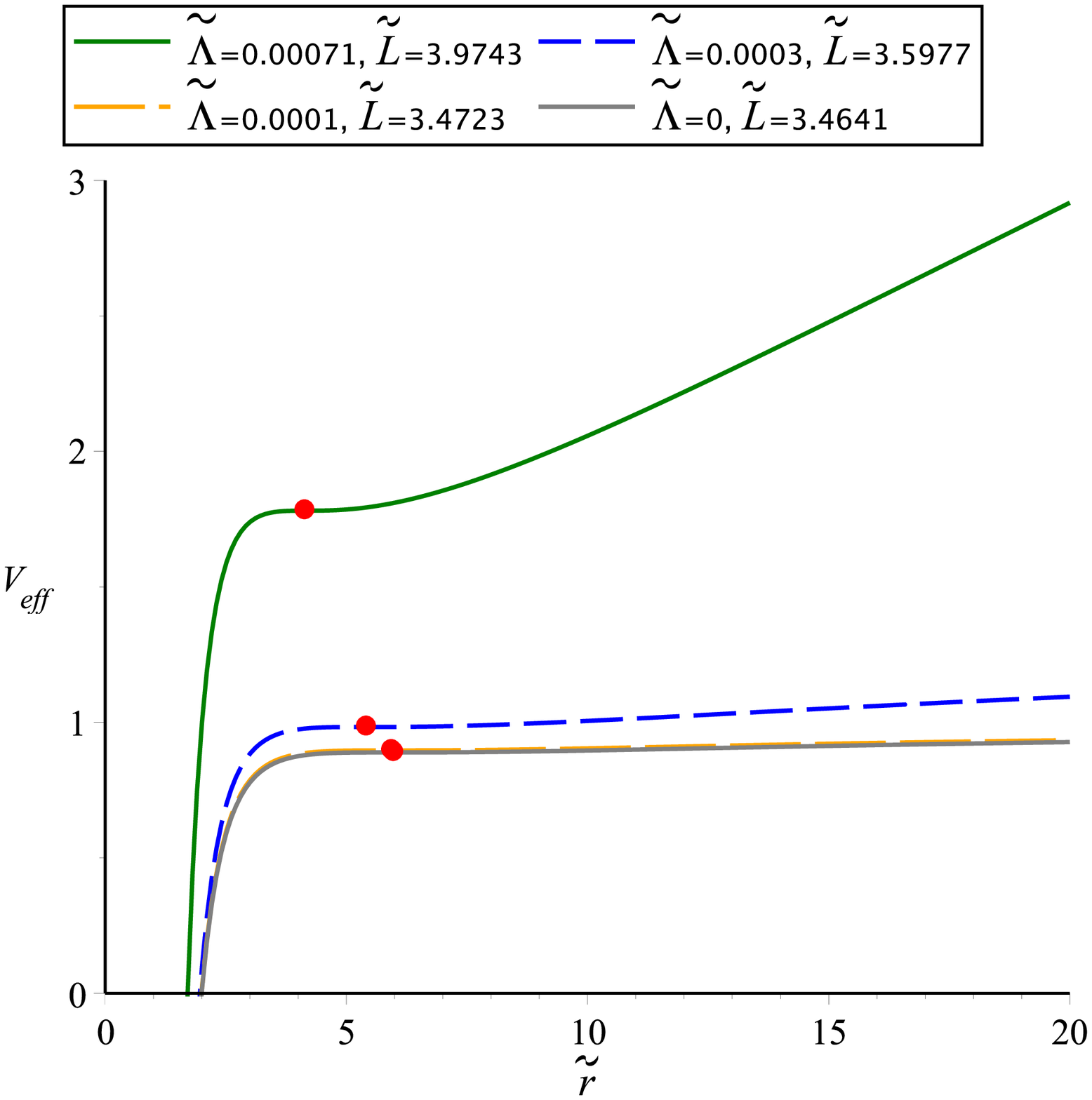}
	}
	\caption{Effective potential for different values of
		$ \tilde{\Lambda} $ and $ \tilde{L} $ of black hole in  f(R) gravity. The dots indicate
		the location of the innermost stable circular orbit.}
	\label{Pic:V}
\end{figure}

\begin{figure}[h]
	\centering
	\subfigure{
		\includegraphics[width=0.6\textwidth]{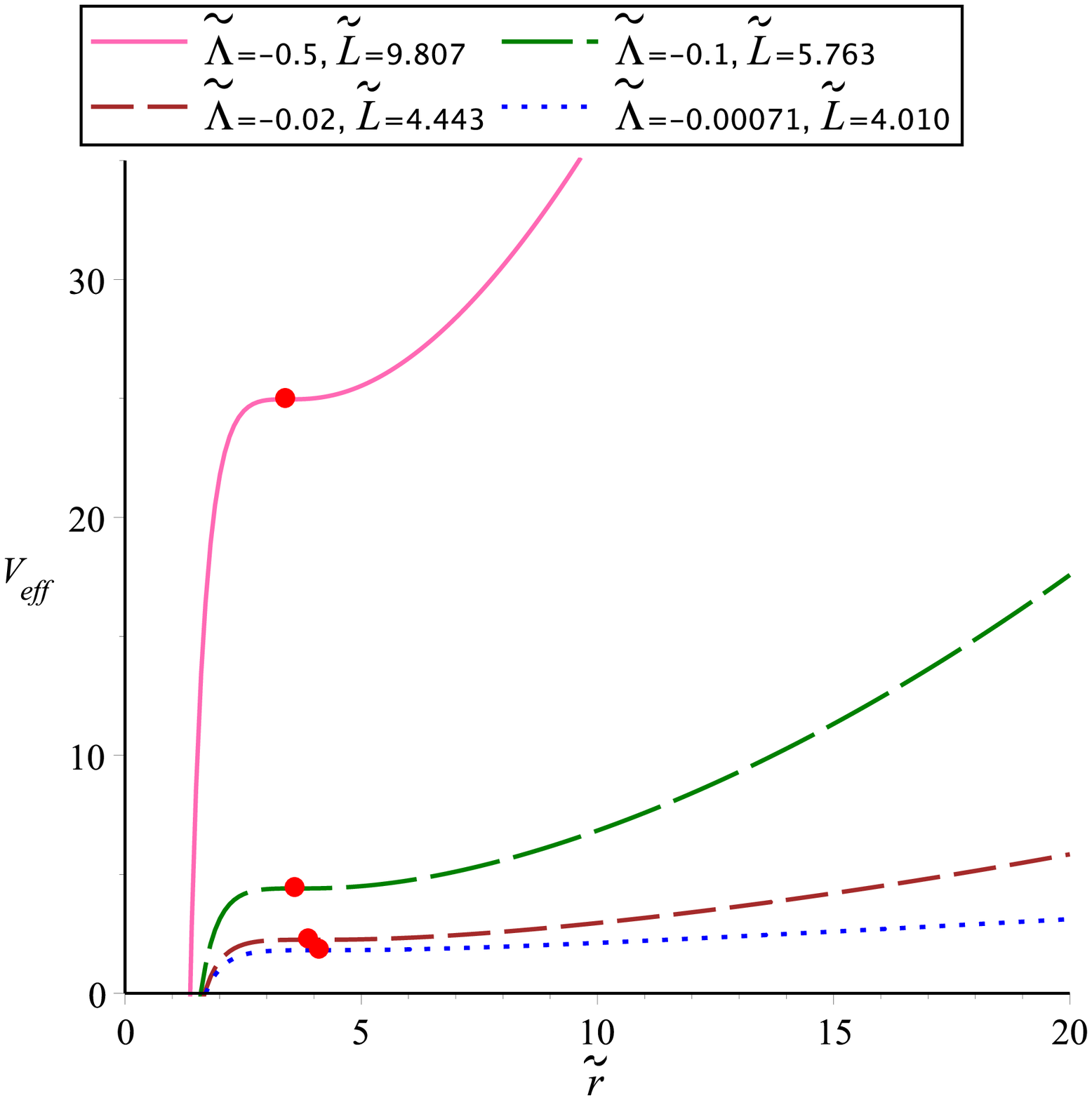}
	}
\caption{Effective potential for different values of $ \tilde{\Lambda}<0 $ and $ \tilde{L} $ 
of black hole in  f(R) gravity. The dots 
 indicate
the location of the innermost stable circular orbit.}
	\label{Pic:Vm1}
\end{figure}

  The accretion disk appears due to the particles revolving in a circular orbit around a compact object such as a black hole. 
The specific energy $\tilde E$, the specific angular momentum $\tilde L$ and the angular velocity $\Omega$, of the particles moving in circular orbits can be calculated, respectively,  by
\begin{eqnarray}
\tilde{E}&=&\dfrac{g_{tt}}{\sqrt{g_{tt}-g_{\phi\phi}\Omega^{2}}}=- \frac{{\sqrt 2 }}{3}\frac{{\left( {\tilde\Lambda {\mkern 1mu} {\tilde r^3} - 3{\mkern 1mu} \tilde \beta {\mkern 1mu} {\tilde r^2} - 3{\mkern 1mu} \tilde r + 6} \right)}}{  \sqrt {  \tilde r({\tilde \beta {\mkern 1mu} {\tilde r^2} + 2{\mkern 1mu}\tilde  r - 6})  }     },\label{E}\\ 
 \tilde{L}&=&\dfrac{g_{\phi\phi}\Omega}{\sqrt{g_{tt}-g_{\phi\phi}\Omega^{2}}}={\mkern 1mu}  {\mkern 1mu}  {\tilde {r}\sqrt { - \frac{{2{\mkern 1mu} \tilde \Lambda {\mkern 1mu} {\tilde r^3} - 3{\mkern 1mu}\tilde  \beta {\mkern 1mu} {\tilde r^2} - 6}}{ 3(  {\tilde \beta {\mkern 1mu} {\tilde r^2} + 2{\mkern 1mu}\tilde  r - 6} ) }} } \label{L},\\
{\Omega}&=&\sqrt{\dfrac{g_{tt,r}}{g_{\phi\phi,r}}}= {\mkern 1mu} \sqrt { - \frac{{2{\mkern 1mu}\tilde  \Lambda {\mkern 1mu} {\tilde r^3} - 3{\mkern 1mu}\tilde  \beta {\mkern 1mu} {\tilde r^2} - 6}}{ 6\tilde r^3 }} \label{Om}.
\end{eqnarray}
Using Eqs. (\ref{V}) and (\ref{L}), we have
\begin{equation}\label{V2}
\frac{{{d^2}{V_{eff}}}}{{{dr^2}}} =  - \frac{2}{3}{\mkern 1mu} \frac{{(3{\mkern 1mu}\tilde  \Lambda {\mkern 1mu}\tilde  \beta {\mkern 1mu} {\tilde r^5} - 3{\mkern 1mu} {\tilde \beta ^2}{\tilde r^4} + 8{\mkern 1mu}\tilde  \Lambda {\mkern 1mu} {\tilde r^4} - 30{\mkern 1mu}\tilde  \Lambda {\mkern 1mu} {\tilde r^3} - 9{\mkern 1mu}\tilde  \beta {\mkern 1mu} {\tilde r^3} + 36{\mkern 1mu}\tilde  \beta {\mkern 1mu} {\tilde r^2} - 6{\mkern 1mu} \tilde r + 36)}}{{{\tilde r^3}\left( {\tilde \beta {\mkern 1mu} {\tilde r^2} + 2{\mkern 1mu}\tilde  r - 6} \right)}}.
\end{equation}
Note that   the necessary condition for existence of innermost circular orbits is  $ \dfrac{d^{2}V_{eff}}{dr^{2}} = 0 $  and the signs of $ \dfrac{d^{2}V_{eff}}{dr^{2}} $ show  the stability of orbits. Thus, by equating Eq. (\ref{V2}) to zero, we obtain the $ \tilde{\Lambda} $ as a function of the radial coordinate of the circular orbits as following:
\begin{equation}
 \tilde{\Lambda}=3{\mkern 1mu} \frac{{{\tilde \beta ^2}{\tilde r^4} + 3{\mkern 1mu} \tilde \beta {\mkern 1mu} {\tilde r^3} - 12{\mkern 1mu}\tilde  \beta {\mkern 1mu} {\tilde r^2} + 2{\mkern 1mu}\tilde  r - 12}}{{{\tilde r^3}\left( {3{\mkern 1mu} \tilde \beta {\mkern 1mu} {\tilde r^2} + 8{\mkern 1mu} \tilde r - 30} \right)}}.
\end{equation}
Plot of the this equation is shown in Fig. \ref{Pic:LambdaBeta}. From the plot, it is evident that the values of $ \tilde{\Lambda} $ is an increasing function of  $\tilde \beta$.
Also, plot of $ \tilde{\Lambda} $ as a function of the radial co-ordinate of the event horizon and of $ \tilde{\Lambda} $ as a function of the radial coordinate of the innermost stable circular orbits is shown in Fig. \ref{Pic:LambdaBeta12}.
Moreover, plots of effective potential (Eq.\ref{Vtilde}) with the location of innermost circular orbit are shown in Figs. (\ref{Pic:V}, \ref{Pic:Vm1}) for $ \tilde{\Lambda}>0 $ and $ \tilde{\Lambda}<0 $ respectively. The dots indicate the location of the innermost stable circular orbits. 
Locations of the event and cosmological horizons, and of the innermost and outermost stable circular orbits in a static black hole in $f(R)$ gravity are shown in Tables. (\ref{tab:Lambdabeta}, \ref{tab:Lambdabetam}) for $ \tilde{\Lambda}>0 $ and $\tilde{\Lambda}<0 $ respectively.

\begin{table}[ht]
		\centering
	\begin{tabular}{|c| c| c| c| c| c|}
		\hline\hline
		&   $\tilde{\beta} $ & \thead{Event horizon\\  radius}  & \thead{Cosmological\\ horizon radius} & \thead{Innermost stable\\  circular orbit\\ radius} & \thead{Outermost stable\\ circular orbit\\ radius}  \\
		\hline\hline
		& 0 & 2 & $ \times $ &
		6
		& $ \times $
		\\ \cline{2-6}
		& 0.001 & 1.9960 & $ \times $ &
		5.9001
		& $ \times $
		\\  \cline{2-6}
		$ \tilde{\Lambda}=0 $ & 0.01 &  1.9615 & $ \times $ & 
		5.3675
		& $ \times $
		\\ \cline{2-6}
		& 0.1  & 1.7082 & $ \times $ &
		4.14
		& $ \times $
		\\ \cline{2-6}
		& 0.99 & 1.0033 & $ \times $ & 
		2.4393
		& $ \times $
		\\ \hline
		& 0 & 2.0024 & 56.7077 & 6.2425 & 12.2499
		\\ \cline{2-6}
		& 0.001 & 1.9984 & 58.4294 & 6.0963 & 13.7866
		\\  \cline{2-6}
		$ \tilde{\Lambda}=0.0003 $ & 0.01 & 1.9367 & 76.0250 & 5.4355 & 38.0848
		\\ \cline{2-6}
		& 0.1  & 1.7093 & 342.9949 & 4.1518 & 336.3436
		\\ \cline{2-6}
		& 0.99 & 1.0034 & 3301.0091 & 2.4395 & 3300.3357
		\\ \hline
		& 0 & 2.0057 & 36.4862 & 7.4038 & 7.6016
		\\ \cline{2-6}
		& 0.001 & 2.0016 & 37.2180 & 6.6190 & 8.9474
		\\  \cline{2-6}
		$ \tilde{\Lambda}=0.00071 $ & 0.01 & 1.9667 & 44.3863 & 5.5474 & 17.9842
		\\ \cline{2-6}
		& 0.1  & 1.7108 & 150.1032 & 4.1598 & 143.5071
		\\ \cline{2-6}
		& 0.99 & 1.0035 & 1395.3741 & 2.4397 & 1394.7007
		\\ \hline
	\end{tabular}
	\caption{Location of the event and cosmological horizons, and of the innermost and outermost stable circular orbits in a static black hole in f(R) gravity.}\label{tab:Lambdabeta}
\end{table}

\begin{table}[ht]
		\centering
	\begin{tabular}{| c| c| c| c|}
		\hline\hline
		&   $\tilde{\Lambda} $ & \thead{Event horizon radius}  & \thead{Innermost stable 
		circular orbit radius}  \\
		\hline\hline
		& -0.00071 & 1.994 & 5.669  
		\\ \cline{2-4}
		& -0.003 & 1.977 & 5.192
	\\  \cline{2-4}
		$ \tilde{\beta}=0 $ & -0.02 &  1.869 & 4.429 
		\\ \cline{2-4}
		&- 0.1  & 1.595 & 3.988
    	\\ \cline{2-4}
		& -0.5 & 1.180 & 3.809 
		\\ \hline
		& -0.00071 & 1.956 & 5.237 
		\\ \cline{2-4}
		&-0.003 & 1.940 & 4.968 
		\\  \cline{2-4}
		$ \tilde{\beta}=0.01 $ & -0.02 & 1.841 & 4.363 
		\\ \cline{2-4}
		& -0.1  & 1.580 & 3.943 
		\\ \cline{2-4}
		& -0.5 & 1.175 & 3.761
		\\ \hline
		& -0.00071 & 1.706 &  4.133 
		\\ \cline{2-4}
		& -0.003 & 1.697 & 4.094 
		\\  \cline{2-4}
		$ \tilde{\beta}=0.1 $ & -0.02 & 1.642 & 3.904 
		\\ \cline{2-4}
		& -0.1  & 1.468 & 3.619
		\\ \cline{2-4}
		& -0.5 & 1.137 & 3.420 
			\\ \hline
		& -0.00071 & 1.003  & 2.439 
		\\ \cline{2-4}
		& -0.003 & 1.002 & 2.437 
		\\  \cline{2-4}
		$ \tilde{\beta}=0.99 $ & -0.02 & 0.9967 & 2.427 
		\\ \cline{2-4}
		& -0.1  & 0.9723 & 2.387
		\\ \cline{2-4}
		& -0.5 &  0.8832 & 2.283
			\\ \hline
	\end{tabular}
	\caption{Location of the event horizon, and of the innermost stable circular orbit in a 
	static black hole in f(R) gravity.}\label{tab:Lambdabetam}
\end{table}

The flux of the radiant energy over the disk can be calculated from following relation 
\cite{pag,pag1}
\begin{equation}\label{F}
F(r)=\dfrac{-\dot{M}_{0}}{4 \pi \sqrt{-g}}\dfrac{\Omega_{,r}}{(\tilde{E}-\Omega \tilde{L} 
)^{2}}\int^{r}_{r_{ms}}(\tilde{E}-\Omega \tilde{L} )  \tilde{L}_{,r} dr .
\end{equation}
 which should follow
Stefan-Boltzmann law when the disk is supposed to be in thermal equilibrium. 
\clearpage  
\section{Concluding remarks}\label{section6}
Let us summarize our work here.  In order to study  the thin accretion disc for the static spherically symmetric black hole in $f(R)$ gravity, we have given preliminary idea about  the metric and the action for such system. We have computed the horizon of the black hole by equating metric function to zero. Furthermore, we have expressed metric function in terms of dimensionless parameters $\tilde{r}, \tilde{\Lambda}$ and $\tilde{\beta}$. In order to discuss
the horizons, we have plotted   $\tilde{\Lambda}$ with respect to $\tilde{r}$ for 
different values of $\tilde{\beta}$. For $\tilde\beta=0.01$ case,  we have found an inner black hole event horizon and an
outer cosmological horizon for
 $\tilde{\Lambda}\in (0,  0.04069)$. In this case,  
 only single black hole event horizon is seen for $\tilde{\Lambda} \leq 0$.
  The event and cosmological
horizons collapse for $\tilde{\Lambda} = 0.04069$ and for higher values naked singularities occur and hence, no black
holes are possible. Consequently,  the trajectories should be studied for $\tilde{\Lambda}\in (-\infty,  0.04069]$.
Similarly, we have studied   $\tilde\beta=0.1$ and $\tilde\beta=0.99$ cases. 

In order to discuss a thin accretion disk around  the spherically symmetric black hole in $f(R)$ gravity, we have computed the specific energy, the specific angular momentum and the
angular velocity  of the particles moving in circular orbits.
The equation of motion and effective potential are also computed.
From  the necessary condition (i.e. the double derivative of effective potential with respect to radial co-ordinate  must be zero) 
for existence of innermost circular orbits of accretion disc,  
we have computed an expression for dimensionless cosmological constant $ \tilde{\Lambda} $.   
Here, we have done comparative analysis
of $ \tilde{\Lambda} $ as a function of $\tilde{r}$ corresponding to different values of 
dimensionless real parameter $\tilde{\beta}$ and 
found that the value of  dimensionless cosmological constant  increases with larger values of 
real parameter. Furthermore, we have discussed
  $ \tilde{\Lambda} $ as a function of the radial co-ordinate of the event horizon and  
  $ \tilde{\Lambda} $ as a function of the radial coordinate of the innermost stable circular 
  orbits.
Finally, we  have studied  the effective potential  with respect to $\tilde{r} $
for $\tilde{\Lambda}\geq 0$ and different $\tilde{L}$ where the locations of   the innermost stable circular orbits are indicated.  
The effective potential  with respect to $\tilde{r} $
for $\tilde{\Lambda} < 0$ and different $\tilde{L}$ are also studied, where the locations of   the innermost stable circular orbits are indicated.
We have provided two tables, one specifying   
locations of the event and cosmological horizons, and of the innermost and outermost stable 
circular orbits in a static black hole in $f(R)$ gravity  and other specifying locations of the event horizon, 
and of the innermost stable circular orbit in a static black hole in $f(R)$ gravity.

\end{document}